\documentstyle[preprint,psfig,aps,floats,tighten]{revtex} % for preprint mode
\def\Journal#1#2#3#4{{#1} {\bf #2}, #3 (#4)}

\def\NIMA{Nucl. Instrum. Methods A}
\def\NPB{Nucl. Phys. B}
\def\PRL{Phys. Rev. Lett.}
\def\PRD{Phys. Rev. D}

\newcommand{\met}{\mbox{$\not\!\!E_T$}}
\newcommand{\metc}{\mbox{$\not\!\!E_T^{\rm cal}$}}

\begin{document}

\preprint{Fermilab-Pub-97/059-E}
\title{Direct Measurement of the Top Quark Mass}

\author{
\centerline{The D\O\ Collaboration\thanks{Authors listed on the following page.
            \hfill\break 
            Submitted to Physical Review Letters.}}
}

\address{
\centerline{Fermi National Accelerator Laboratory, Batavia, Illinois 60510}
}

\date{March 10, 1997}

\maketitle

\begin{abstract}
We measure the top quark mass $m_t$ using $t{\bar t}$ pairs produced in the 
D\O\ detector by $\sqrt{s}$ = 1.8 TeV $p {\bar p}$ collisions in a 
125~pb$^{-1}$ exposure at the Fermilab Tevatron.  We make a two constraint fit 
to $m_t$ in $t{\bar t} \rightarrow bW^+ \, {\bar b}W^-$ final states with one 
$W$ decaying to $q {\bar q}$ and the other to $e \nu$ or $\mu \nu$.  Events 
are binned in fit mass {\it versus} a measure of probability for events to be 
signal rather than background.  Likelihood fits to the data yield 
$m_t = 173.3 \pm 5.6 \; ({\rm stat}) \pm 6.2 \; ({\rm syst}) \; {\rm GeV/c}^2$.
\end{abstract}

\vskip 1cm

\begin{center}
%\small
%% names begin here
S.~Abachi,$^{14}$
B.~Abbott,$^{28}$
M.~Abolins,$^{25}$
B.S.~Acharya,$^{43}$
I.~Adam,$^{12}$
D.L.~Adams,$^{37}$
M.~Adams,$^{17}$
S.~Ahn,$^{14}$
H.~Aihara,$^{22}$
G.A.~Alves,$^{10}$
% ...
\end{center}
\vskip 0.50cm
\normalsize
\centerline{(D\O\ Collaboration)}

%\vskip 0.50cm
\vfill\eject
\small
\it
\centerline{$^{1}$Universidad de los Andes, Bogot\'{a}, Colombia}
\centerline{$^{2}$University of Arizona, Tucson, Arizona 85721}
\centerline{$^{3}$Boston University, Boston, Massachusetts 02215}
\centerline{$^{4}$Brookhaven National Laboratory, Upton, New York 11973}
\centerline{$^{5}$Brown University, Providence, Rhode Island 02912}
\centerline{$^{6}$Universidad de Buenos Aires, Buenos Aires, Argentina}
\centerline{$^{7}$University of California, Davis, California 95616}
\centerline{$^{8}$University of California, Irvine, California 92697}
\centerline{$^{9}$University of California, Riverside, California 92521}
\centerline{$^{10}$LAFEX, Centro Brasileiro de Pesquisas F{\'\i}sicas,
                   Rio de Janeiro, Brazil}
\centerline{$^{11}$CINVESTAV, Mexico City, Mexico}
\centerline{$^{12}$Columbia University, New York, New York 10027}
\centerline{$^{13}$Delhi University, Delhi, India 110007}
\centerline{$^{14}$Fermi National Accelerator Laboratory, Batavia,
                   Illinois 60510}
\centerline{$^{15}$Florida State University, Tallahassee, Florida 32306}
\centerline{$^{16}$University of Hawaii, Honolulu, Hawaii 96822}
\centerline{$^{17}$University of Illinois at Chicago, Chicago, Illinois 60607}
\centerline{$^{18}$Indiana University, Bloomington, Indiana 47405}
\centerline{$^{19}$Iowa State University, Ames, Iowa 50011}
\centerline{$^{20}$Korea University, Seoul, Korea}
\centerline{$^{21}$Kyungsung University, Pusan, Korea}
\centerline{$^{22}$Lawrence Berkeley National Laboratory and University of
                   California, Berkeley, California 94720}
\centerline{$^{23}$University of Maryland, College Park, Maryland 20742}
\centerline{$^{24}$University of Michigan, Ann Arbor, Michigan 48109}
\centerline{$^{25}$Michigan State University, East Lansing, Michigan 48824}
\centerline{$^{26}$Moscow State University, Moscow, Russia}
\centerline{$^{27}$University of Nebraska, Lincoln, Nebraska 68588}
\centerline{$^{28}$New York University, New York, New York 10003}
\centerline{$^{29}$Northeastern University, Boston, Massachusetts 02115}
\centerline{$^{30}$Northern Illinois University, DeKalb, Illinois 60115}
\centerline{$^{31}$Northwestern University, Evanston, Illinois 60208}
\centerline{$^{32}$University of Notre Dame, Notre Dame, Indiana 46556}
\centerline{$^{33}$University of Oklahoma, Norman, Oklahoma 73019}
\centerline{$^{34}$University of Panjab, Chandigarh 16-00-14, India}
\centerline{$^{35}$Institute for High Energy Physics, 142-284 
                   Protvino, Russia}
\centerline{$^{36}$Purdue University, West Lafayette, Indiana 47907}
\centerline{$^{37}$Rice University, Houston, Texas 77005}
\centerline{$^{38}$Universidade Estadual do Rio de Janeiro, Brazil}
\centerline{$^{39}$University of Rochester, Rochester, New York 14627}
\centerline{$^{40}$CEA, DAPNIA/Service de Physique des Particules, CE-SACLAY,
                   Gif-sur-Yvette, France}
\centerline{$^{41}$Seoul National University, Seoul, Korea}
\centerline{$^{42}$State University of New York, Stony Brook, New York 11794}
\centerline{$^{43}$Tata Institute of Fundamental Research,
                   Colaba, Mumbai 400005, India}
\centerline{$^{44}$University of Texas, Arlington, Texas 76019}
\centerline{$^{45}$Texas A\&M University, College Station, Texas 77843}
\normalsize

\vfill\eject
%\twocolumn
The top quark has a large mass $m_t$ that can be determined to greater 
fractional precision than is possible for the lighter quarks, which decay 
after they form hadrons.  Since $m_t$ is large, it controls the strength of 
quark-loop corrections to tree-level relations among electroweak parameters.  
If these parameters and $m_t$ are measured precisely, the Standard Model 
Higgs boson mass can be constrained.
  
Direct measurements of $m_t$ have been published as part of the initial 
observations~\cite{discov} of $t {\bar t}$ production in $\sqrt{s} = 1.8$ 
TeV $p {\bar p}$ collisions.  At present, the best accuracy in $m_t$ is 
achieved for lepton + jets ($\ell$+jets) final states in which one $W$ boson 
(from $t \rightarrow bW$) decays to $e \nu$ or $\mu \nu$ and the other $W$ 
decays to a $q {\bar q}$ pair that forms jets.  We report a measurement of 
$m_t$ in the $\ell$+jets channel using the $\approx$125~pb$^{-1}$ exposure 
of the D\O\ detector during the 1992--96 Fermilab Tevatron runs.  Since 
Ref.~\cite{discov} appeared, our data sample has doubled, and for a fixed 
sample size our error on $m_t$ has halved.
       
The D\O\ detector and our basic methods for triggering, reconstructing events, 
and identifying particles are described elsewhere~\cite{d0prnim}.  Recent 
advances include enhanced triggering and reconstruction efficiency for 
$\mu$+jets events, due in part to better use of calorimeter data.  As a 
signature of $W \rightarrow \ell \nu$, we require missing energy transverse 
to the beam ($\met$) $>$ 20 GeV, and one isolated $e$ or $\mu$ ($\ell$) with 
$E_T^\ell>20$ GeV and pseudorapidity $|\eta_e|<2$ or $|\eta_{\mu}|<1.7$.  
We also demand $\metc > 25$ (20) GeV for $e$+jets ($\mu$+jets) events, where 
$\metc$ is $\met$ measured only in the calorimeter.  As signatures of the 
$q {\bar q}$ from $W$ decay and the $b$ and ${\bar b}$ from $t$ and ${\bar t}$ 
decay, we require $\ge$4 jets reconstructed with cones of half-angle $\Delta 
{\cal R} \equiv (\Delta \phi^2 +  \Delta\eta^2)^{1/2} = 0.5$, having $E_T>15$ 
GeV and $|\eta|<2$.

\begin{figure}[b]
%\centerline{\psfig{figure=prl7-fig1-bw.eps,width=3.23in}}
%\centerline{\psfig{figure=prl7-fig1.eps,width=3.23in}}
%\centerline{\psfig{figure=prl7-fig1.eps,width=4.3in}}
\centerline{\psfig{figure=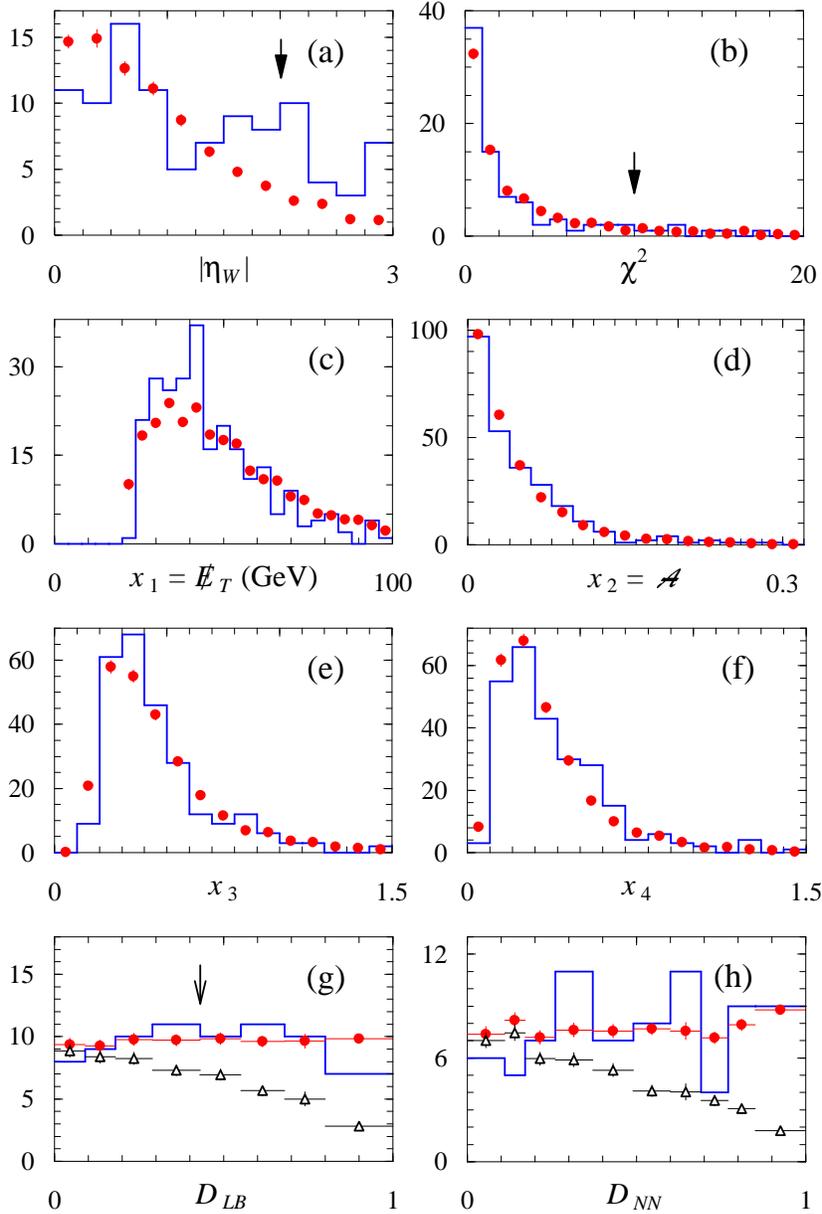,width=4.3in}}
%\vskip 0.1cm
\vskip 0.5cm
\caption{
\label{fig1}
Events per bin {\it vs.}~event selection variables defined in the text, 
plotted for (a--b, g--h) top quark mass analysis samples, and (c--f) $W$+3 
jet control samples.  Histograms are data, filled circles are expected top + 
background mixture, and open triangles are expected background only.  Solid 
arrows in (a--b) show cuts applied to all events; the open arrow in (g) 
illustrates the LB cut.  The nonuniform bin widths in (g--h) are chosen to 
yield uniform bin populations.
}
\end{figure}

Within $\Delta {\cal R} = 0.5$ of a jet axis, additional muons ($\mu$ tags) 
satisfying $p_T^\mu > 4$ GeV/$c$ and $|\eta_\mu| < 1.7$ arise mainly from $b$ 
and $c$ quark semileptonic decay.  These occur in $\approx$20\% of 
$t {\bar t}$ events but only $\approx$2\% of background events~\cite{d0prnim}. 
In untagged events, to suppress background we require $E_T^L$ ($\equiv 
|E_T^\ell| + |\met|$) $> 60$ GeV and $|\eta_W| < 2$ for the $W \rightarrow 
\ell \nu$.  The latter cut, exhibited in Fig.~\ref{fig1}(a), reduces the 
difference in $\eta_{W}$ distributions between data and Monte Carlo (MC) 
simulated background.  We use the {\sc herwig} MC~\cite{herwig} to simulate 
top signal, and the {\sc vecbos} MC~\cite{vecbos} (with {\sc herwig} 
fragmentation of partons into jets) to simulate (but not to normalize) the 
dominant $W$+multijet background.  The $\approx$20\% of background events from 
non-$W$ sources are modeled by multijet data that barely fail the lepton 
identification criteria.

To each event passing the above cuts, we make a two constraint (2C) kinematic
 fit~\cite{snyder} to the $t {\bar t} \rightarrow \ell$+jets hypothesis by 
minimizing a $\chi^2 = ({\bf v} - {\bf v}^*)^T G ({\bf v} - {\bf v}^*)$, 
where ${\bf v}$ (${\bf v}^*$) is the vector of measured (fit) variables and 
$G^{-1}$ is its error matrix.  Both reconstructed $W$ masses are constrained 
to equal the $W$ pole mass, and the same fit mass $m_{\rm fit}$ is assigned to 
both the $t$ and ${\bar t}$ quarks.  If the event contains $>$4 accepted jets, 
only the four jets with highest $E_T$ are used.  In $\approx$50\% of MC top 
events, these jets correspond to the $b$, ${\bar b}$, $q$, and ${\bar q}$.   
With (without) a $\mu$ tag in the event, there are 6 (12) possible fit 
assignments of these jets to the quarks, each having two solutions to the 
$\nu$ longitudinal momentum $p^\nu_z$.  We use $m_{\rm fit}$ only from the 
permutation with lowest $\chi^2$, the correct choice for $\approx$20\% of MC 
top events.  Because of the ambiguities, $m_{\rm fit}$ is not the same as 
$m_t$, though they are strongly correlated.  Our best estimate of $m_t$ is 
obtained from the best match between MC samples and the data.
            
From the 90-event distribution shown in Fig.~\ref{fig1}(b) we select 77 events 
with a 2C fit satisfying $\chi^2 < 10$.  Of these, 5 are $\mu$ tagged and 
$\approx$65\% are background.  Further separation of signal and background 
events is based on four kinematic variables {\bf x} $\equiv$ 
\{$x_1,x_2,x_3,x_4$\} chosen to have small correlation with $m_{\rm fit}$.  
On average, all are larger for MC top events than for background events, 
selected to have the same $\langle m_{\rm fit} \rangle$ as the top 
events~\cite{pbarp96}.  The simpler variables are $x_1 \equiv \met$ and $x_2 
\equiv {\cal A}$, where aplanarity ${\cal A}$ is ${3 \over 2} \times$ the 
least eigenvalue of the normalized laboratory momentum tensor of the jets and 
the $W$ boson.  The third variable $x_3 \equiv H_{T2} / H_z$ measures the 
event's centrality, where $H_z$ is the sum of $|p_z|$ of $\ell$, $\nu$, and 
the jets, and $H_{T2}$ is the sum of all jet $|E_T|$ except the highest.  
Finally, $x_4 \equiv \Delta {\cal R}_{jj}^{\rm min} E_T^{\rm min} / E_T^L$ 
measures the extent to which jets are clustered together, where $\Delta 
{\cal R}_{jj}^{\rm min}$ is the minimum $\Delta {\cal R}$ of the six pairs of 
four jets, and $E_T^{\rm min}$ is the smaller jet $E_T$ from the minimum 
$\Delta {\cal R}$ pair.  As shown for the background dominated $W$+3 jet 
sample in Fig.~\ref{fig1}(c--f), $x_1$--$x_4$ are reasonably well modeled by 
MC; this is true also for the $W$+2 jet and top mass samples (not shown).    

We bin events in a two-dimensional array with abscissa $m_{\rm fit}$ and 
ordinate $D({\bf x})$, where $D$ is a multivariate discriminant.  To show that 
our results are robust, we use two methods for which the definition of $D$, 
the granularity with which it is binned, and the additional requirements are 
different.  In our ``low bias'' (LB) method, we first parametrize 
${\cal L}_i(x_i) \equiv s_i(x_i)/b_i(x_i)$, where $s_i$ and $b_i$ are the top 
signal and background densities in each variable, integrating over the others. 
We form the log likelihood $\ln{\cal L} \equiv \sum_i \omega_i 
\ln{{\cal L}_i}$, where the weights $\omega_i$ are adjusted slightly away 
from unity to nullify the average correlation (``bias'') of ${\cal L}$ with 
$m_{\rm fit}$, and for each event we set $D_{\rm LB} = {\cal L} / (1 + 
{\cal L})$.  Finally, we divide the ordinate coarsely into signal- and 
background-rich bins according to whether the LB cut is passed.  This cut is 
satisfied if a $\mu$ tag exists; otherwise it is not satisfied if 
$D_{\rm LB} < 0.43$ (Fig.~\ref{fig1}(g)) or if $H_{T2} < 90$ GeV.

\begin{figure}[b]
%\centerline{\psfig{figure=prl62-fig2-bw.eps,width=3.22in}}
%\centerline{\psfig{figure=prl62-fig2.eps,width=3.22in}}
%\centerline{\psfig{figure=prl62-fig2.eps,width=4in}}
\centerline{\psfig{figure=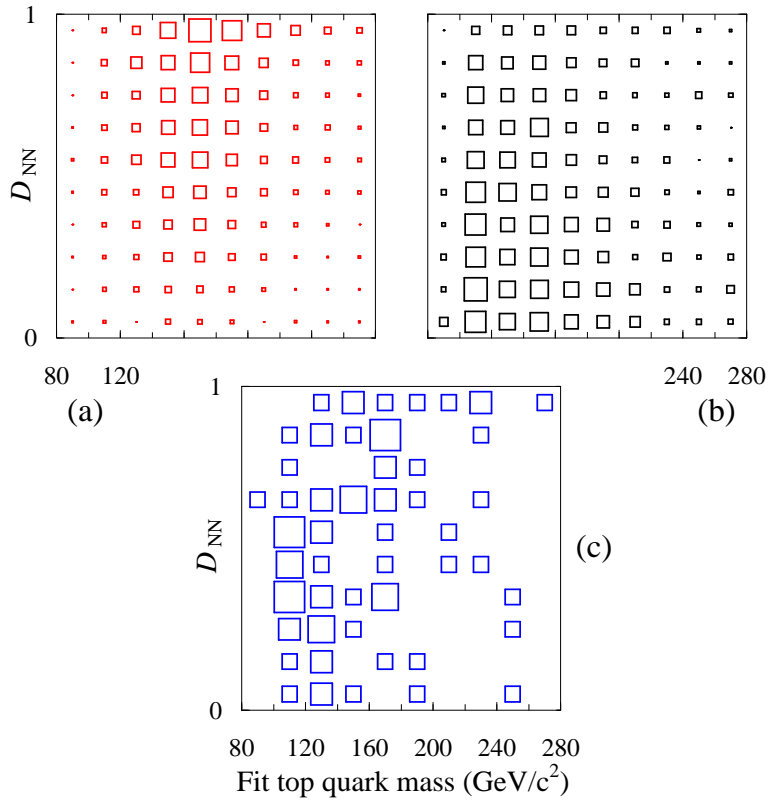,width=4in}}
%\vskip 0.1cm
\vskip 0.5cm
\caption{
\label{fig2}
Events per bin ($\propto$ areas of boxes) {\it vs.}~$D_{\rm NN}$ (ordinate) 
and $m_{\rm fit}$ (abscissa) for (a) expected 172 GeV/c$^2$ top signal, 
(b) expected background, and (c) data.  $D_{\rm NN}$ is binned as in 
Fig.~\ref{fig1}(h).
}
\end{figure}

\begin{figure}[b]
%\centerline{\psfig{figure=prl62-fig3-nofill-bw.eps,width=3.375in}}
%\centerline{\psfig{figure=prl62-fig3.eps,width=3.375in}}
%\centerline{\psfig{figure=prl62-fig3.eps,width=4.2in}}
\centerline{\psfig{figure=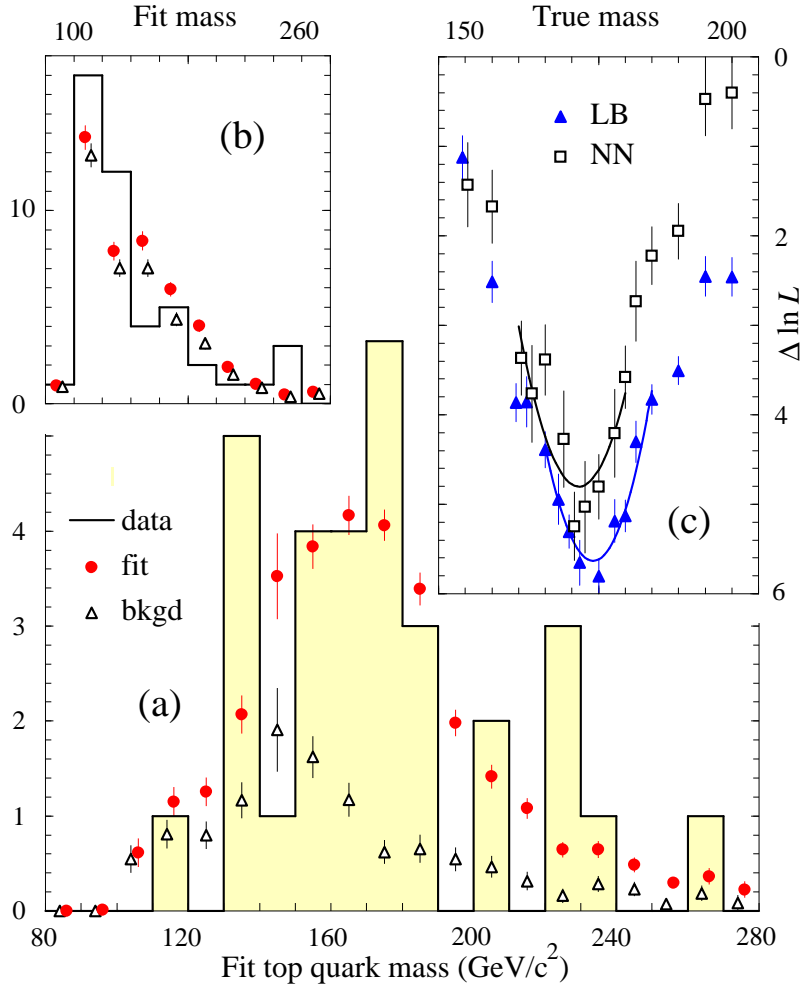,width=4.2in}}
%\vskip 0.1cm
\vskip 0.5cm
\caption{
\label{fig3}
(a--b) Events per bin {\it vs.}~$m_{\rm fit}$ for events (a) passing or (b) 
failing the LB cut.  Histograms are data, filled circles are the predicted 
mixture of top and background, and open triangles are predicted background 
only.  The circles and triangles are the average of the LB and NN fit 
predictions, which differ by $<$10\%. (c) Log of arbitrarily normalized 
likelihood $L$ {\it vs.}~true top quark mass $m_t$ for the LB (filled 
triangles) and NN (open squares) fits, with errors due to finite top MC 
statistics.  The curves are quadratic fits to the lowest point and its 8 
nearest neighbors.  In MC studies, 7\% (27\%) of simulated experiments yield 
a smaller LB (NN) maximum likelihood.
}
\end{figure}

Our neural network (NN) method is sensitive to the correlations among the 
$x_i$ as well as to their individual densities.  We use a three layer 
feed-forward NN with 4 input nodes fed by {\bf x}, 5 hidden nodes, and 1 
output node, trained on samples of top signal (background) with density 
$s({\bf x})$ ($b({\bf x})$)~\cite{neural}.  For a given event, the network 
output $D_{\rm NN}$ approximates the ratio $s({\bf x}) / (s({\bf x}) + 
b({\bf x}))$.  We divide the ordinate finely into ten bins in $D_{\rm NN}$, 
independent of $H_{T2}$ or $\mu$ tagging.  Figure \ref{fig1}(g--h) shows 
that $D_{\rm LB}$ and $D_{\rm NN}$ are distributed as predicted and provide 
comparable discrimination, as we expect when the $\omega_i$ are close to 
unity and the ${\cal L}_i$ are not strongly correlated.  Figure~\ref{fig2} 
exhibits the arrays for the NN method.  Little correlation between 
$D_{\rm NN}$ and $m_{\rm fit}$ is evident in the expected signal or background 
distributions, which are distinct; the data clearly reveal contributions from 
both sources.  Figure~\ref{fig3} shows the distributions of $m_{\rm fit}$ for 
data (a) passing and (b) failing the LB cut.

To each $m_t$ for which we have generated MC, we assign a likelihood $L$ which 
assumes that all samples obey Poisson statistics.  Bayesian 
integration~\cite{harry} over possible true signal and background populations 
in each bin yields
%\begin{eqnarray}
%L(m_t,n_s,n_b) \ &&= \ \prod_{i=1}^M \, \sum_{j=0}^{n_i} 
%  \left( \begin{array}{c} n_{si}+j \\ j \end{array} \right)   
%  \left( \begin{array}{c} n_{bi}+k \\ k \end{array} \right)
%  \nonumber \\
%  \ &&\; \ 
%  \times p^j_s (1+p_s)^{-n_{si}-j-1}   
%  \,
%  p^k_b (1+p_b)^{-n_{bi}-k-1}
%  \; , \nonumber
%\end{eqnarray}
\begin{displaymath}
L(m_t,n_s,n_b) = \prod_{i=1}^M \, \sum_{j=0}^{n_i} 
  \left( \begin{array}{c} n_{si}+j \\ j \end{array} \right)   
  \left( \begin{array}{c} n_{bi}+k \\ k \end{array} \right)
  p^j_s (1+p_s)^{-n_{si}-j-1} \,   
  p^k_b (1+p_b)^{-n_{bi}-k-1}
  \; ,
\end{displaymath}
where $n_s$ ($n_b$) is the expected number of signal (background) events in 
the data; $n_i$, $n_{si}$, and $n_{bi}$ are the actual number of data, MC 
signal, and MC background events in bin $i$; $k \equiv n_i - j$; $p_{s,b} 
\equiv n_{s,b} / (M + \sum_i{n_{si,bi}})$; and $M= 40$ (200) bins for the LB 
(NN) methods.  Maximizing $L$ for each $m_t$ gives the best estimates 
$n^*_s(m_t)$ and $n^*_b(m_t)$ for $n_s$ and $n_b$.  Figure~\ref{fig3}(c) 
displays $\ln{L(m_t, n^*_s(m_t), n^*_b(m_t))}$ {\it vs.}~$m_t$, where the 
curves determine the best fit $m_t$ and its statistical error $\sigma_m$.

\begin{table}[t]
%\vskip 0.5cm
\caption{
\label{tab1}
Results of fits to data and MC events.  Fits to data yield values and errors 
$\sigma$(stat)~for $m_t$, $n_s$, and $n_b$ (described in the text).  
Systematic errors are combined in quadrature.  The resulting $m_t$ and its 
statistical error $\sigma_m$ are the combined LB and NN values.  Fits to MC 
use ensembles of 10,000 simulated experiments composed of top + background, 
with $m_t$, $\langle n_s \rangle$, and $\langle n_b \rangle$ as listed.  They 
yield a mean result $\langle m_t \rangle$, a mean statistical error $\langle 
\sigma_m \rangle$, and a range $\pm \delta m$ within which 68\% of the results 
fall.  Using the LB (NN) method, 6\% (25\%) of the simulated experiments 
produce a $\sigma_m$ which is smaller than we obtain.  For an ``accurate 
subset'' of the MC ensembles with mean $\sigma_m/m_t$ that matches our value, 
$\delta m$ is smaller.     
}
%\vskip 0.1cm
\vskip 0.5cm
%\centerline{\psfig{figure=prl7-tab1-bw.eps,width=3.4in}}
%\centerline{\psfig{figure=prl7-tab1.eps,width=3.4in}}
%centerline{\psfig{figure=prl7-tab1.eps,width=4.3in}}
\centerline{\psfig{figure=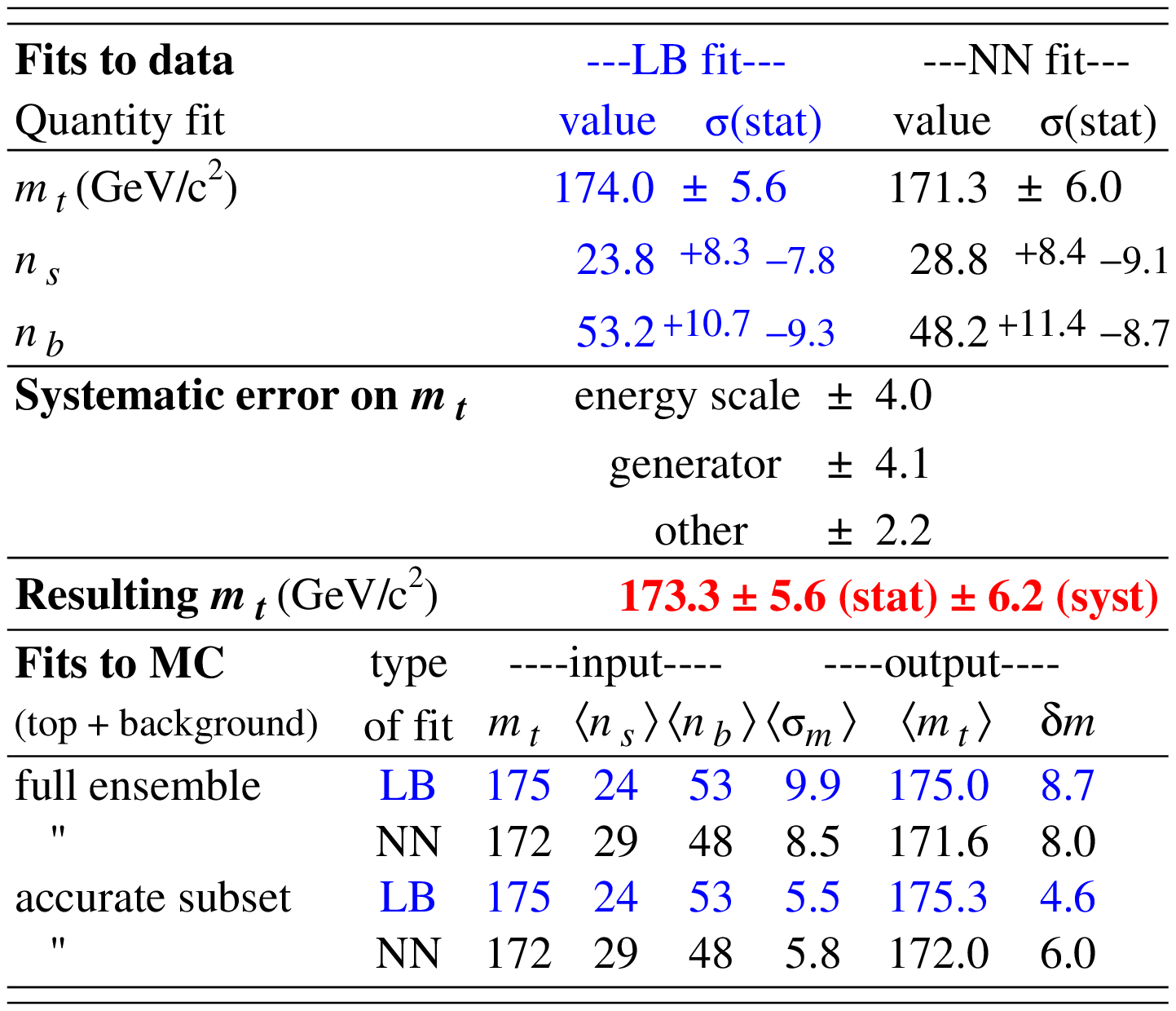,width=4.3in}}
\end{table}

Table~\ref{tab1} presents the fit results, which are consistent with 
Ref.~\cite{discov} and with recent reports~\cite{serban}.  The LB and NN 
results $m_t^{\rm LB}$ and $m_t^{\rm NN}$ are mutually consistent; in 21\% of 
MC experiments they are further apart.  Nevertheless we include half of 
$m_t^{\rm LB} - m_t^{\rm NN}$ in the systematic error.  To obtain our result, 
shown in Table~\ref{tab1}, we combine $m_t^{\rm LB}$ and $m_t^{\rm NN}$ 
allowing for their ($88 \pm 4$)\% correlation (determined by MC experiments).  
Figures~\ref{fig3}(a--b) show that this result represents the data well.  
From the MC experiments summarized in Table~\ref{tab1} we measure the 
interval $\pm \delta m$ within which 68\% of the MC estimates fall.  For the 
full ensemble, $\delta m$ is larger than $\sigma_m$ from our data.  However, 
for ``accurate subsets'' of the ensemble for which the average $\sigma_m/m_t$ 
is the same as we observe, $\delta m$ is close to $\sigma_m$~\cite{stochastic}.
           
A principal systematic error in $m_t$ arises from uncertainty in the jet 
energy scale, which is calibrated in three steps.  In step 1, applied before 
events are selected, the summed energy $E_{\rm jet}$ of particles emitted 
within the jet cone is related~\cite{kehoe} to the measured energy $E_{\rm m}$ 
by $E_{\rm jet} = (E_{\rm m} - O)/ R(1-S)$.  Here the calorimeter response 
$R$ is calibrated using $Z \rightarrow ee$ decays and $E_T$ balance in 
$\gamma$+jet events, the fractional shower leakage $S$ out of the jet cone is 
set by test beam data, and the energy offset $O$ due to noise and the 
underlying event is determined using events with multiple interactions.  Steps 
2 and 3 are applied only to jet energies used to find $m_{\rm fit}$.  In 
step 2, top MC is used to correct $E_{\rm jet}$ to the parton energy in both 
data and MC.  This sharpens the resolution in $m_{\rm fit}$.  Step 3 is a 
final adjustment based on more detailed study of $\gamma$+jet events in data 
and MC, particularly focused on the dependence of the $E_T$ balance upon 
$\eta$ of the jet.  We assign a jet-scale error of $\pm$(2.5\% + 0.5 GeV) 
based on the internal consistency of step 3, on variations of the $\gamma$+jet 
cuts and the model for the underlying event, and on an independent check of 
the $E_T$ balance in $Z$+jet events.  This leads to an error on $m_t$ of 
$\pm$4.0 GeV/c$^2$.

We estimate the uncertainties in modeling of QCD by substituting the {\sc 
isajet} MC generator~\cite{isajet} for {\sc herwig}, independently for top MC 
and for {\sc vecbos} fragmentation, and by changing the {\sc vecbos} QCD scale 
from jet $\langle p_T \rangle^2$ to $M_W^2$.  The resulting systematic error
due to the generator is $\pm$4.1 GeV/c$^2$.  Other effects including noise, 
multiple $p{\bar p}$ interactions, and differences in fits to $\ln{L}$ 
contribute $\pm$2.2 GeV/c$^2$.   All systematic errors (Table~\ref{tab1}) sum 
in quadrature to $\pm$6.2 GeV/c$^2$.  Therefore our direct measurement of the 
top quark mass is
$m_t = 173.3 \pm 5.6 \; \rm{(stat)} \pm 6.2 \; \rm{(syst)}$ GeV/c$^2$.

We thank the staffs at Fermilab and the collaborating institutions for their 
contributions to this work, and the Department of Energy and National Science 
Foundation (U.S.A.), Commissariat  \` a L'Energie Atomique (France), State 
Committee for Science and Technology and Ministry for Atomic Energy (Russia), 
CNPq (Brazil), Departments of Atomic Energy and Science and Education (India), 
Colciencias (Colombia), CONACyT (Mexico), Ministry of Education and KOSEF 
(Korea), CONICET and UBACyT (Argentina), and the A.P.~Sloan Foundation for 
support.


\begin{references}
\bibitem{stochastic} We have varied our analysis procedures (e.g.~the 
binning of $D$) in ways which have little systematic effect on MC results.  
From data we observe little change in $m_t$, together with variations in 
$\sigma_m$ which are of the same order as those of $\delta m$ in 
Table~\ref{tab1}.  We interpret the variations in $\sigma_m$ as stochastic 
effects to which the MC studies in Table~\ref{tab1} are relevant.
\end{references}
\end{document}